\documentclass[manuscript]{aastex}
\usepackage{amssymb}
\usepackage{longtable}
\usepackage{array}

\usepackage{amsmath}

\slugcomment{Not to appear in Nonlearned J., 45.}
\shorttitle{Solar Spatial Magnetic Field}
\shortauthors{Liu D.}
\begin{document}
\title{Relationship between Magnetic Field Properties and an X-class Flare
in Active Region NOAA 9077}
%%\author{D. Liu\altaffilmark{1}}
%\author{S. Liu, D. Liu}
%\altaffiltext{1}{Liao Ning University, Shenyang, China\\}
%\email{liudian@lnu.edu.cn}
\author{S. Liu\altaffilmark{1,2}, D. Liu\altaffilmark{2}}

\affil{
$^{1}$National Astronomical Observatory, Chinese Academy of
Sciences,
        Beijing, China\\
$^{2}$Liao Ning University,
        Shenyang, China}

\email{liud@lnu.edu.cn}
\email{lius@nao.cas.cn}

\altaffiltext{1}{key Laboratory of Solar Activity}
\begin{abstract}
The magnetic field plays a key role in producing solar flares, so that
the investigation on the relationship between the magnetic field properties
and flares is significant. In this paper, based on the magnetic field
extrapolated from the photospheric vector magnetograms of the active region NOAA 9077
obtained at Huairou Solar Observing Station, the magnetic field parameters
including the height of field lines, force-free factor, free magnetic energy and inclination angle
were studied with respect to an X-class flare in this region. We found that
the magnetic field lines became lower and the ratio of number of closed field lines to those
of open field lines increased after the flare. The force-free factor ($\alpha$) attained
a large value before the flare and then decreased after the flare for the closed
field lines, while the open field lines showed the opposite tendency.
Free energy reach to maximum before flare, then decrease after flare.
The magnetic inclination angles showed opposite change trends after the flare
for closed and open field lines. Therefore, we may conclude that non-potential energy released by flare
mostly contained in the closed magnetic field lines.

\end{abstract}


\section{INTRODUCTION}
Solar active region dominated by magnetic field that count for most energetic solar activities such as flare, corona mass ejection and filaments eruption
\citep{kra82, pri03, wie14}.
Hence, it becomes inevitable to study the relations between magnetic field and flare. Some physical parameters deduced from magnetic field become
the targets to investigate the properties of magnetic field and flare \citep{mci90, lv93, dyy01, nin02, liu08}. The above researches based on
magnetic field observed on the photosphere, the only layer where magnetic field can be measured with high accuracy, as a result the fully understands
about magnetic field and flare are limited due to the absence of spatial magnetic field observations on the sun.

The active region space magnetic field can be obtained by magnetic extrapolation assuming field above active region is force-free, which can be regarded a reasonable
assumption \citep{aly89} for low $\beta$ (the ratio of gas pressure to magnetic pressure) plasma circumstances, where magnetic field satisfy $(\nabla\times\textbf{\emph{B}}) \times\textbf{\emph{B}}=0$ and $\nabla\cdot\textbf{\emph{B}}=0$. Recently, lots of force-free extrapolation methods
are developed and applied to extrapolated chromosphere and corona field  \citep{wu90, mic94, ama97, sak81, yan00, whe00, wie04, son06, he08, liu11a, liu11b}.
So the tentatively studies about the fully spatial magnetic field and flare become possible basing on magnetic extrapolation \citep{yan01,wie05,liu11a,jiang13,liu14,ama14}. Then, the studies about magnetic parameters can been deduced from spatial magnetic field become available,
which may service to the knowledge of magnetic field and understanding of solar activities.

Solar flares are typical eruption events with its energy comes from magnetic field free energy.
It is certain that there exist the changes of magnetic field before and after flares, for kinetic and
thermal energy of flare are converted from the magnetic energy. The previous studies have shown some possible changes
of magnetic field related parameters before and after solar flare. For example, the increases of transverse magnetic
field were found after flare for some active regions \citep{wan05, su11, wan12},
there are evident decreases of free energy after the solar flares \citep{sun12}.
When magnetic field extrapolations are applied to observations, the spatial magnetic field parameters and their
relationships with flare are expected to be investigated further.

Systematic observations of the vector magnetic field of active region 9077 from 11 to 15 July were carried out
at Huairou Solar Observing Station (HSOS), National Astronomical Observatories
of Chinese Academy of Sciences (NAOC). On 14 July 2000, a giant solar flare with an X-ray
importance of X5.7 exploded near the disk center in active region 9077.
It is one of the greatest solar events in that solar cycle, and the flare caused major terrestrial effects.
Based on these high quality vector magnetic field with famous flare event and advanced field extrapolation, it give us a chance to
study the spatial magnetic field and it relations with this giant flare. The main targets here is to study spatial field lines distributions,
magnetic inclination angles and force-free factor, those magnetic parameters are basically related non-potentiality of active region magnetic field.

This paper is arranged as follows: Section 2 describes the observations data and extrapolation method;
in Section 3 the results are given; discussions and conclusions will be given in Section 4.

\section{OBSERVATIONS DATA AND EXTRAPOLATION METHOD}
The photosphere vector magnetic field used as boundary for extrapolation in this paper
is observed by the Solar Magnetic Field Telescope (SMFT) \citep{ai86} at HSOS.
With SMFT, which is a 35cm solar telescope, the photospheric vector magnetic field can be measured with Fe I $\lambda$ 5324.19 \AA.
Offset shifted -0.075 \AA~relative to center of Fe I $\lambda$ 5324.19 \AA~used to measure the longitudinal
magnetic field and the center of line is applied to measure the transverse components.
The vector magnetic field is produced by calibrating Stokes parameters ($Q$, $U$
and $V$). $V$ is the difference of the left and right circularly polarized images, $Q$ and $U$ are the differences between
two orthogonal linearly polarized images for different azimuthal directions. As for vector magnetic field calibration, series
works are carried \citep{wan96,su04}, here calibration coefficients of $C_{L}$ and $C_{T}$ are 8381 G and 6790 G are used under
the weak-field approximation \citep{jef89, jef91}:
\begin{equation}
B_{L}=C_{L}V ,\\
\end{equation}
\begin{equation}
B_{T}=C_{T}(Q^{2}+U^{2})^{1/4}, \\
\end{equation}
\begin{equation}
\theta=arctan(\dfrac{B_{L}}{B_{T}}),\\
\end{equation}
\begin{equation}
\phi=\dfrac{1}{2}arctan(\dfrac{U}{Q}),\\
\end{equation}
Here, $\theta$ is the
inclination between the vector magnetic field and the direction
normal to the solar surface and $\phi$ is the field azimuth.
Through data processing of SMFT vector magnetic field, the spatial
resolution of observational data is about ~2$''$/pixel $\times$
~2$''$/pixel and noise levels of vector magnetograms are 5
G and 50 G for longitudinal and transverse components,
respectively. The observation with the field of view 5.23$'$ $\times$ 3.63$'$ during this period.
The method of minimum energy are employed to resolve $180^{\circ}$ ambiguity,
in which both the divergence of the magnetic field ($\nabla\cdot\textbf{\emph{B}}$) and
the electric current density ($\textbf{\emph{J}}=\nabla\times\textbf{\emph{B}}$)
are minimized simultaneously using a simulated annealing algorithm \citep{mat94,mat06}.
%\citep{wan94,wang97,wang01,mat06}.
Lastly, the vector magnetic field projected to local heliographic coordinate is used as boundary conditions
for magnetic field extrapolation.

The optimization method is used in this paper to extrapolate spatial
magnetic field of active region.
This method is presented by \citep{whe00}
and developed by \citep{wie04}, it consists by minimizing both
normalized Lorentz force and the divergence of the field, which is given by the
following function,
\begin{equation}
\label{opit} L = \int_{V}\omega(x,y,z)[B^{-2}|(\nabla \times
\textbf{B}\times \textbf{B}) |^{2}+|\nabla\cdot
\textbf{B}|^{2}]d^{3}x,
\end{equation}
where $\omega(x,y,z)$ is regarded as a space weighting function.
It is undoubt that the force-free conditions can be satisfied when $L$ is
equal to zero. This method involves minimizing $L$ by optimizing the
solution function $\textbf{B}(x,t)$ through states that are
increasingly force- and divergence-free, where $t$ is an artificial
time-like parameter introduced in this algorithm.

\section{RESULTS}
\begin{figure}

%%%
  % \centerline{\includegraphics[width=1\textwidth,clip=]{./te-boundary.ps}}
   \centerline{\includegraphics[width=0.8\textwidth,clip=]{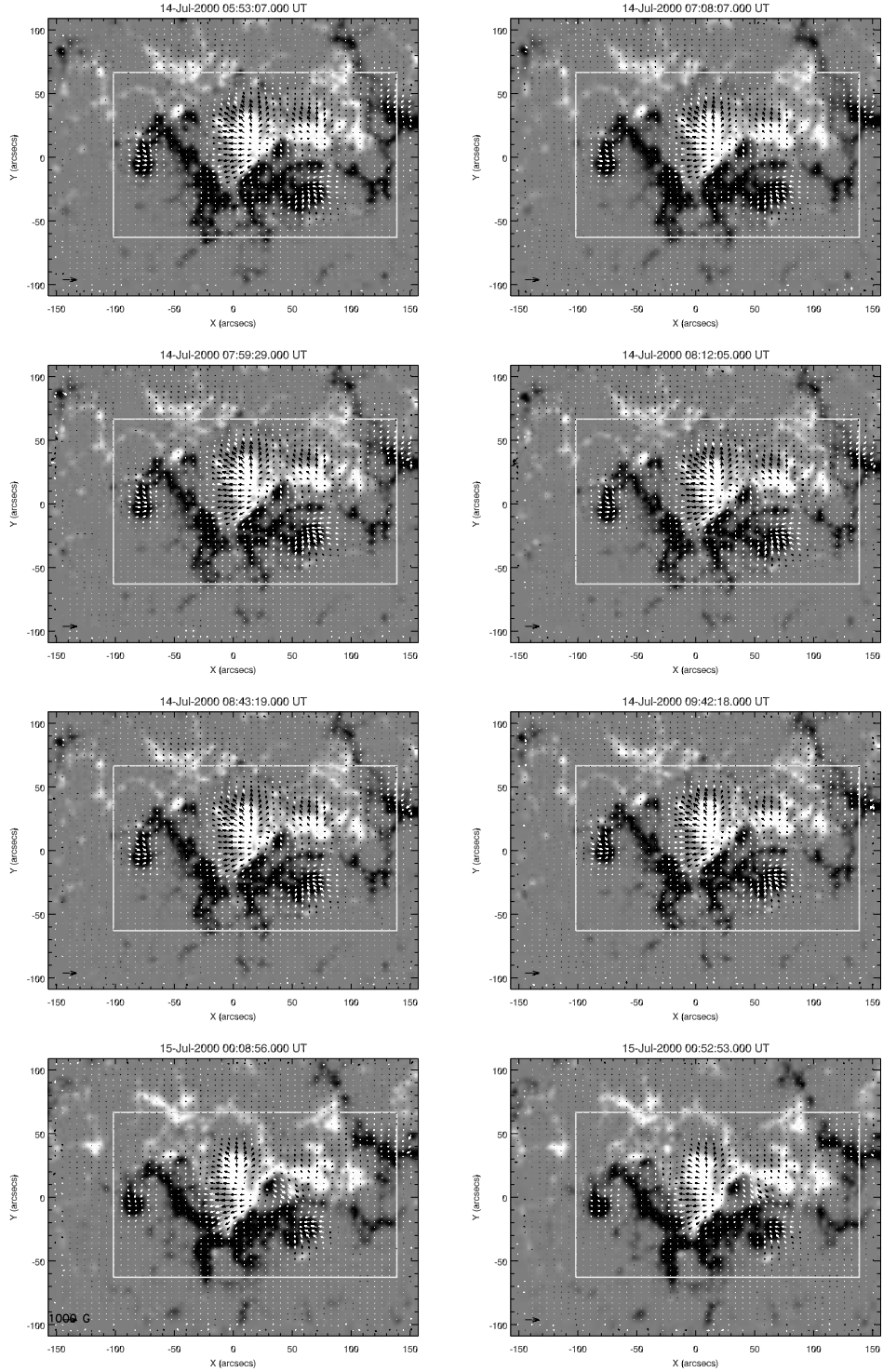}}

   \caption{The magnetograms of the observing active region 9077 with the field of view  5.23$'$ $\times$ 3.63$'$ before and after the flare, times are labelled in each panel respectively. Arrow directions and its length indicate the directions and
strength of the transverse field.} \label{Fig1}
\end{figure}

\begin{figure}

%%%
  % \centerline{\includegraphics[width=1\textwidth,clip=]{./te-boundary.ps}}
   \centerline{\includegraphics[width=0.8\textwidth,clip=]{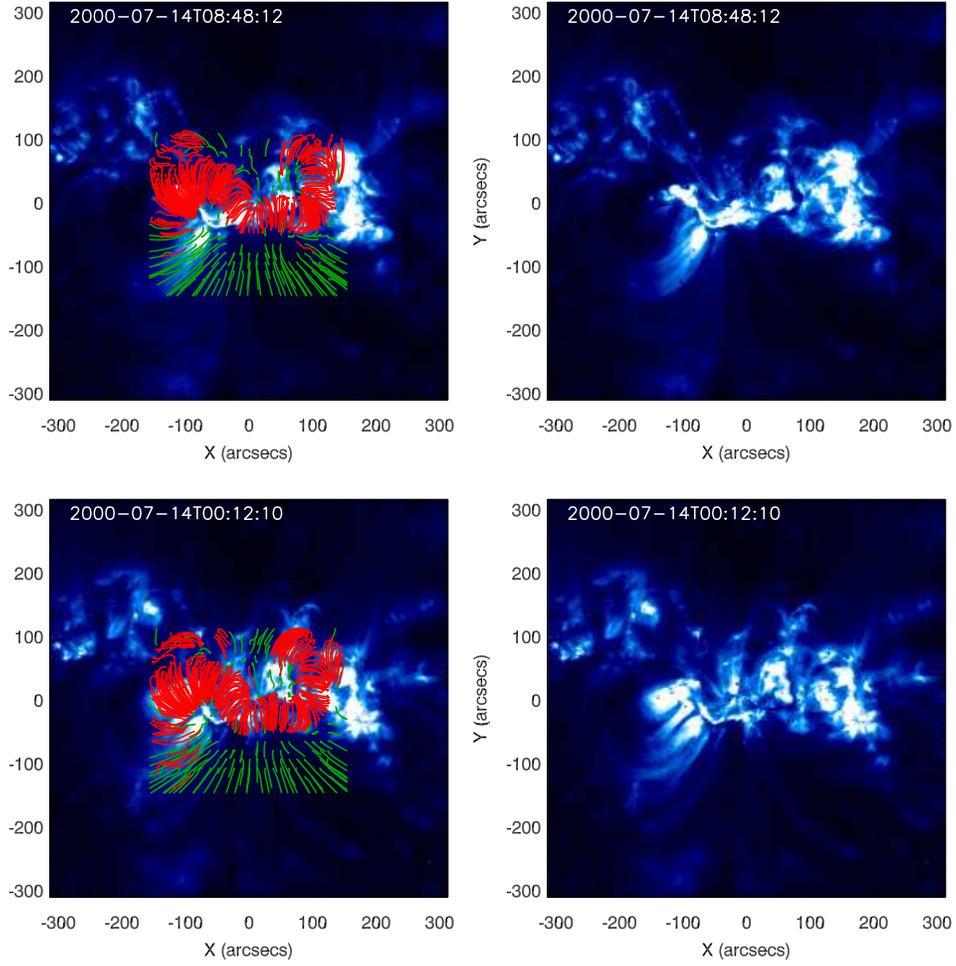}}

   \caption{The extrapolated field lines distributions overlaid on the image of EIT/SOHO 195 \AA, the red lines are closed and the green ones are open.} \label{Fig_eit}
\end{figure}

\begin{figure}

%%%
   %\centerline{\includegraphics[width=1\textwidth,clip=]{./lines_distr.ps}}
   \centerline{\includegraphics[width=1\textwidth,clip=]{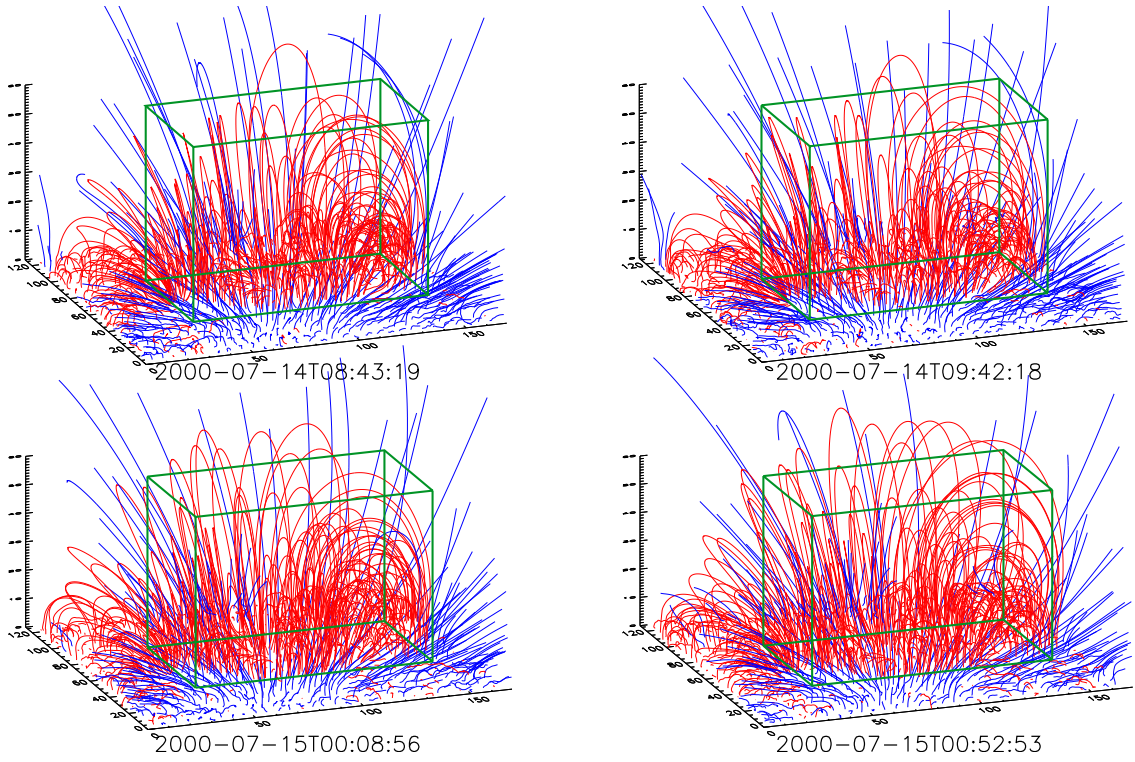}}

   \caption{The spatial magnetic field line distributions extrapolated from the magnetic field given in Fig 1.
The red lines are closed and the blue represent open lines.} \label{Fig2}
\end{figure}

Figure \ref{Fig1} shows the vector magnetic fields of the active region observed 9077 before and after the X5.7 flare that erupted at 10:03 on 14 July 2000,
the six magnetograms in top three rows were observed before this flare, and the bottom after it.
The individual observed time is labelled in each magnetogram.
Here, the field of view is 5.23$'$ $\times$ 3.63$'$, the arrow directions and its length indicate the directions and
strength of the transverse fields. The active region can be regarded as a super one \citep{chen11} with evident
$\beta\gamma\delta/\beta\gamma\delta$ structures (Maximum area: 1010 $\mu$h),
which was accompanied by an Earth-directed coronal mass ejection (CME) and consequently caused major terrestrial effects .
The white rectangles in each panel is the region where the magnetic parameters were calculated.
Figure \ref{Fig_eit} shows two examples of the field lines overlaid on the images of EIT/SOHO 195 \AA~ channel, where red lines are closed in the extrapolation space while the green
lines are ones escape the space (here the field of view of EIT are larger than that of extrapolated),
 it can indicate that most field lines match coronal loops on the whole, so the researches about spatial magnetic field and
 flare become available basing on extrapolated fields.
Figure \ref{Fig2} shows the spatial magnetic field line distributions extrapolated from the magnetic fields of the magnetograms given
in Figure 1 as boundary conditions, where the red lines are closed field lines and the blue ones represent open lines,
and boxes in green (here the height $\approx$~66 Mm) correspond
to the white rectangles in Figure 1. Here only four cases as examples are shown,
the top two panels observed before flare and the lower two after it.
On the whole, there are more closed field lines after the flare than those before the flare, by which we were motivated
to explore in more detail the information contained in the field extrapolation.

\begin{figure}

%%%
   %\centerline{\includegraphics[width=1\textwidth,clip=]{./line_group_count_bili.ps}}
   \centerline{\includegraphics[width=1\textwidth,clip=]{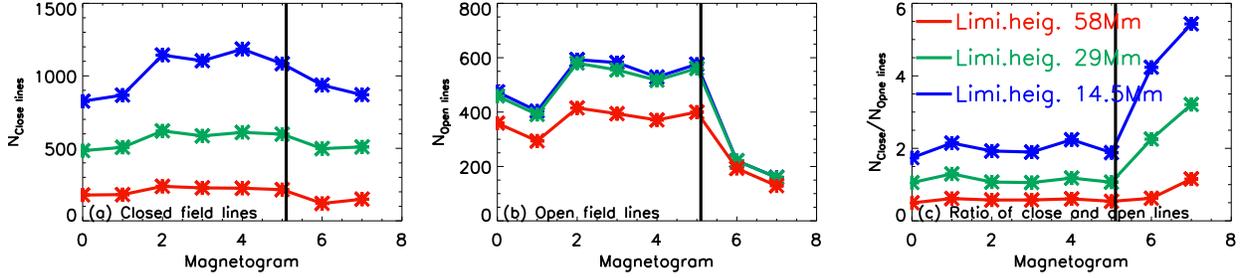}}

   \caption{The evolution of the numbers of closed and open field lines, and the ratio of the closed to open lines shown in (a), (b) and (c).
The blue, green and red curves indicate the field lines  that are higher than 14.5, 29 and 58 Mm, respectively.
Here six/two magnetograms observed before/after the flare are calculated.
In each panel the vertical lines following the data point based on the sixth magnetogram in fiugre \ref{Fig1} indicates the occurrence of the X5.7 flare.} \label{Fig3}
\end{figure}

\begin{figure}

%%%
   %\centerline{\includegraphics[width=1\textwidth,clip=]{./line_heiz.ps}}
    \centerline{\includegraphics[width=1\textwidth,clip=]{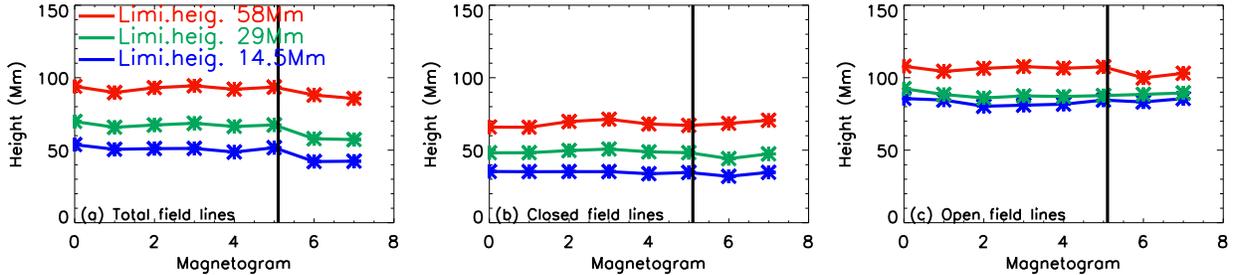}}

   \caption{The evolution of the average heights of total lines,
closed field lines and the open field lines shown in (a), (b) and (c).
The blue, green and red lines indicate the field lines that are higher than 14.5, 29 and 58 Mm are calculated, respectively.
Here six/two magnetograms observed before/after the flare are included and calculated.
In each panel the vertical lines following the data point based on the sixth magnetogram in fiugre \ref{Fig1} indicates the occurrence of the X5.7 flare.} \label{Fig4}
\end{figure}

Figure \ref{Fig3} shows the evolution of the numbers of closed field lines, and open
field lines, and the ratio of closed to open field lines. The calculated regions are labelled by green boxes in Figure 2.
Blue, green and red curve indicate the height ranges of the field lines are higher than 14.5 (10 pixels), 29 (20 pixels) and 58 (40 pixels) Mm, respectively (the limitations of line heights can be regarded as a threshold).
This is done only for relatively larger scale magnetic structures are considered for X5.7 flare, without small scale ones involved temporarily. Six/two magnetograms
(corresponding in Figure 1) observed before/after the flare are included and calculated. In each panel in figure \ref{Fig3} the vertical line following the
sixth data point indicates the occurrence of the X5.7 flare. To avoid the noise interference, the thresholds of of the observed photospheric longitudinal and transverse fields
greater than 20G and 100G respectively are set. In general the closed and open field lines become less after the flare. Additionally, the open field lines decrease faster than closed ones, which results in
a sharp increase of the ratio of the number of closed field lines to that of open lines shown in panel (c) in figure \ref{Fig3}.
Figure \ref{Fig4} shows the evolution of the average heights of the field lines. Same as Figure 3, the calculated heights are higher than 14.5, 29 and 58 Mm
shown by blue, green and red curves, respectively, and the thresholds of 20 and 100 G are set for the observed photospheric longitudinal and transverse fields.
Panel (a), (b) and (c) represent lines of the total field (the sum of closed and open field lines), closed field lines and open field lines, respectively. On the whole, it can be found that after the flare these relative
large scale field lines become lower than those before flare, and this trend is more evident for the lower closed field lines in blue and green shown
in (b) in figure  \ref{Fig4}.

\begin{figure}

%%%
   %\centerline{\includegraphics[width=1\textwidth,clip=]{./line_theta.ps}}
   \centerline{\includegraphics[width=1\textwidth,clip=]{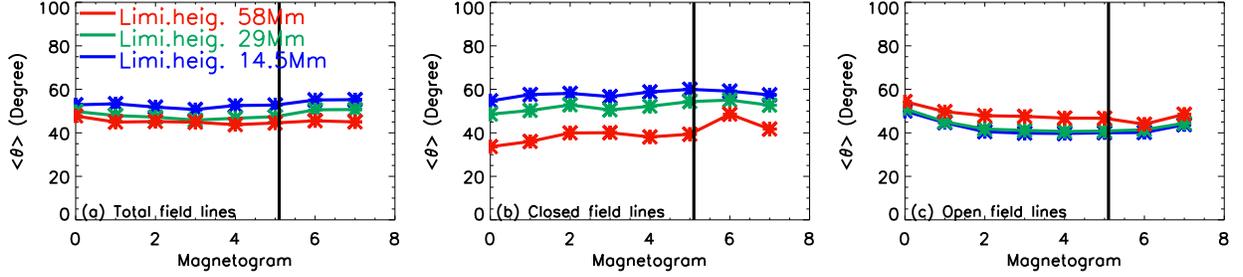}}

   \caption{Same as Fig 4, but (a), (b) and (c) showing the evolution of the averages inclinations angles.} \label{Fig5}
\end{figure}

\begin{figure}

%%%
   %\centerline{\includegraphics[width=1\textwidth,clip=]{./line_alpha.ps}}
     \centerline{\includegraphics[width=1\textwidth,clip=]{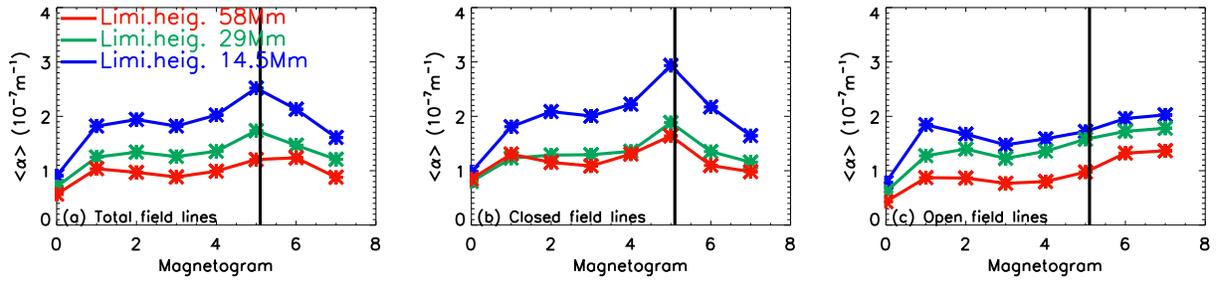}}

   \caption{Same as Fig 4, but (a), (b) and (c) showing the evolution of the averages force free factor ($\alpha$).} \label{Fig6}
\end{figure}

\begin{figure}

%%%
   %\centerline{\includegraphics[width=1\textwidth,clip=]{./line_alpha.ps}}
     \centerline{\includegraphics[width=1\textwidth,clip=]{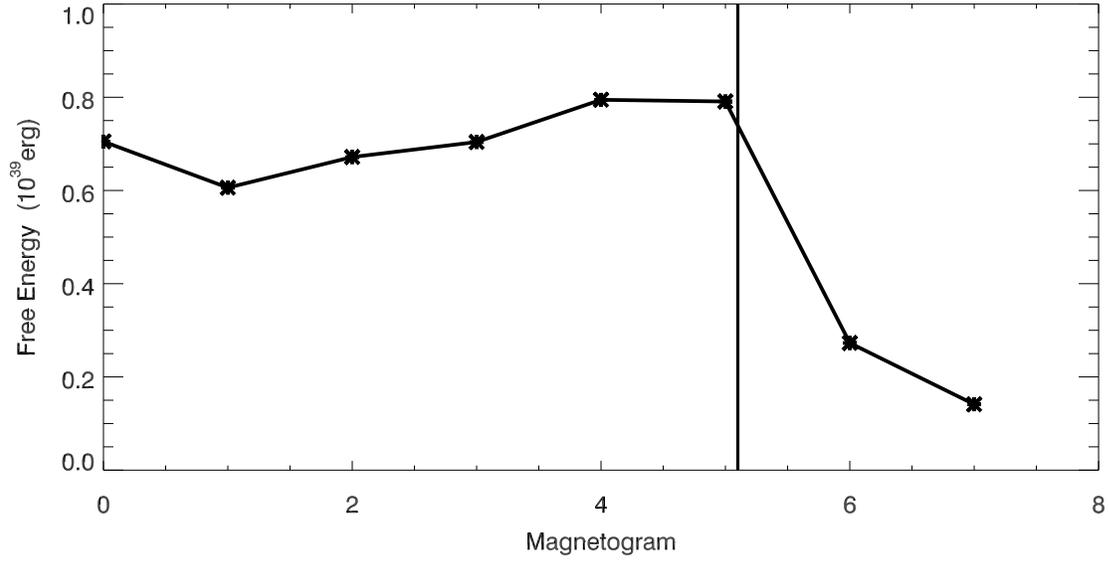}}

   \caption{The evolution of magnetic free energy calculated in the whole space extrapolated.} \label{Fig7}
\end{figure}

Figure \ref{Fig5} (a), (b) and (c) show the evolution of average inclination angles
($atan(B_{z}, B_{t}), Bt=\sqrt{B^{2}_{x}+B^{2}_{y}}$) calculated from total
field lines, closed field lines and open field lines. Same as Figure 3, the calculated heights are higher than 14.5, 29 and 58 Mm
shown by blue, green and red curves, respectively, and the thresholds of 20 and 100 G are set for longitudinal and transverse fields.
It is shown that in (a) for total field lines are increases in inclination after the flare, especially for field lines at low place.
For close field lines (b), at the lower, the inclination decreases after the flare, while at higher height the trend is reverse.
Comparing (b) with (c), the changes of inclination of closed and open field lines are opposite, and this trend is consistent at heights.
However for individual closed and open field lines distributed at different heights the changes in inclination are different,
for example, the changes of red and green(blue) curves in (b) and (c) are different before and after this big flare.
Figure \ref{Fig6} shows the evolution of the average force free factor ($\alpha=\nabla\times\textbf{B}/\textbf{B}=\textbf{B}\cdot\nabla\times\textbf{B}/B^{2}$)
along the total field lines, closed and open field lines.
To take the advantage of the extrapolated spatial magnetic field,
the formula $\alpha=\textbf{B}\cdot\nabla\times\textbf{B}/B^{2}$ effectively avoid the singularity of magnetic components of $B_{x}$, $B_{y}$ and $B_{z}$.
In Figure \ref{Fig6}, after the flare are obvious decrease of $\alpha$ for total field lines and closed field lines,
while for open field lines, the values of $\alpha$ evidently increase after the flare. Additionally, the trend of the changes of $\alpha$
in each panel are consistent at various heights. It is also noted that
throughout the evolution for total field lines and closed field lines the values of $\alpha$ reach its maxima just before this flare occurs.
Free energy (subtract potential magnetic energy from extrapolated NLFF field energy) can been used as a parameter to indicate non-potentiality as $\alpha$, so the evolutions of free energy are shows in the Figure \ref{Fig7},
however it is global parameter it can not been divided by open and closed file lines, here we only give it calculated from the whole volume.
It can been found that the evolutions of free energy are consistent with those of $\alpha$ calculated from the total field lines and closed ones,
it also means that the free energy reach to maximum before flare and decrease evidently after the big flare.

\section{Discussions and conclusions}
The super solar active region NOAA 9077 with the X5.7 flare is used in this paper to investigate the spatial magnetic
field properties, and the changes of the spatial magnetic parameters before and after
the flare are focused in this study. The spatial magnetic parameters are deduced from
the extrapolated field based on the photospheric magnetic field observed and the force free assumption.
\textbf{The reliability of extrapolated field can be seen by the consistences between extrapolated field line and the EIT loops, they are matched
on the whole. The HSOS data adopt linear calibration, so the true magnetic field strength on the sun maybe not very accuracy. However, the topology maybe not depend very strongly on its real strength.}

According to the changes of the magnetic field line distribution with height it may be conclude that the reduction of
the field lines at higher position after the flare, also implies some of them have vanished.
Further as for the decrease of magnetic field lines, the open field lines suffer more than those closed,
which leads to the large increase of the ratio of close field lines to open lines after the flare. This may indicate that there are some evident changes in topologies of magnetic field
due to a part of open field lines possibly becoming closed after the flare. The change of inclination angles in fact is not so remarkable before and after the big flare.
However a rule may be drawn that the direction of inclination changing for closed and open field lines is
mostly opposite , and their trends are different at different heights.
\textbf{However, the changes of line number and inclination angles are very faint, it maybe these two parameters are not much sensitive to magnetic field for this event, so there possible exist a strong and reliable effect.}
Force free factor $\alpha$ deduced from the extrapolated spatial magnetic field shows an interesting pattern
before the flare. Even though $\alpha$ decreases for total field line and closed field lines after flare, as for the open field lines
there is an increase in $\alpha$ in reverse after flare.
Here, force free factor $\alpha$ may be used
to scale the extent of non-potentiality of an active region magnetic field indirectly.
So the changes of $\alpha$ after flare may mean that the most of flare energy comes from closed field lines,
however the open field lines are seriously disturbed by those closed field lines. It should be noted that if some open field lines
become closed ones which can also result in the decrease  of $\alpha$ of closed field lines , however this effect may not be evident.
\textbf{The decreases of $\alpha$ might be regarded as the release of magnetic free energy (the most parts are contained in the close field lines)
to produce a solar flare, and therefore its value reaches maximum just before the flare happens.}

It is conventionally regarded that the solar flare energy comes from magnetic field free energy
and it may be the outcome of magnetic reconnections. Therefore, when a solar flare occurs, the topology of the magnetic field
 will be seriously disturbed consequently. Some previous studies based on
the photosphere magnetic field have been carried out and results obtained exhibit consistent recognition.
In this paper, by means of the advantage of magnetic field extrapolation, the spatial magnetic field is obtained,
then the relationships between the spatial magnetic parameters and the flare are expected to receive further confirmation.

\acknowledgments
%The authors thank the anonymous referee for helpful comments and suggestions.
This work was partly supported by the Grants: 2011CB811401, KLCX2-YW-T04, KJCX2-EW-T07, 2014FY120300, 11373040,
11203036, 11221063, 11178005, 11003025, 11103037, 11103038, 10673016, 10778723 and 11178016, the
Young Researcher Grant of National Astronomical Observations,
Chinese Academy of Sciences, and the Key Laboratory of Solar
Activity National Astronomical Observations, Chinese Academy
of Sciences.

%I:\document-1\hsos-extra\90772\te\te-boundary.pro



\clearpage
\end{document}